\documentclass[twocolumn,showpacs,preprintnumbers,amsmath,amssymb]{revtex4}

\usepackage{graphicx}% Include figure files
\usepackage{dcolumn}% Align table columns on decimal point
\usepackage{bm}% bold math
\usepackage{epsfig,epstopdf}
\usepackage{subfigure}
\usepackage[english]{babel}
\usepackage[latin1]{inputenc}
\begin{document}

\preprint{APS/123-QED}

\title{Collapse of the Quantum Wavefunction}

\author{Håkon Brox}%
%\email{hakonbro@stud.ntnu.no}%
\affiliation{Physics Department, Norwegian University of Science and
  Technology}%Lines break automatically or can be forced with \\

\author{Kåre Olaussen}%
\affiliation{Physics Department, Norwegian University of Science and
  Technology}%Lines break automatically or can be forced with \\

\author{Anh Kiet Nguyen}%
\affiliation{Physics Department, Norwegian University of Science and
  Technology}%Lines break automatically or can be forced with \\

\date{\today}% It is always \today, today,

\begin{abstract}
  We show using a realistic Hamiltonian-type model that definite
  outcomes of quantum measurements may emerge from quantum evolution
  of pure states, i.e quantum dynamics provides a deterministic
  collapse of the wavefunction in a quantum measurement process. The
  relaxation of the wavefunction into a pointer state with classical
  properties is driven by the interaction with an environment. The
  destruction of superpositions, i.e. choosing a preferred attraction
  basin and thereby a preferred pointer state, is caused by a tiny
  nonlinearity in the macroscopic measurement apparatus. In more
  details, we numerically studied the many-body quantum dynamics of a
  closed Universe consisting of a system spin measured by a
  ferromagnet embedded in a spin-glass environment. The nonlinear term
  is the self-induced magnetic field of the ferromagnet. The
  statistics for the outcomes of this quantum measurement process
  depends on the size of the attraction basins in the measurement
  apparatus and are in accordance to Born's rule.
\end{abstract}

\pacs{03.65.Ta, 03.65.Yz, 75.45.+j}

\maketitle

%%% New commands %%%
\newcommand{\eq}{\! = \!}
\newcommand{\keq}{\!\! = \!\!}
\newcommand{\kadd}{\! + \!}

\section{Introduction}
Quantum mechanics~\cite{Bohr,Dirac,neumann} is perhaps the most
successful theory ever developed and provides very accurate
descriptions of Nature. For example the gyromagnetic ratio is
correctly predicted by quantum electrodynamics to twelve significant
figures of accuracy~\cite{QED}. There are at present no experimental
findings that contradict the predictions made by the quantum theory!
In spite of this enormous predicting power, the orthodox quantum
theory~\cite{Bohr,Dirac,neumann} can only provide the probability for
the outcomes of measurements. It can not describe how the complex
quantum wavefunction describing the state of the system actually
evolves into a single classical outcome, found in all measurement
processes. To overcome this problem, the orthodox quantum theory
postulates the discontinuous collapse of the wavefunction upon
measurement, linking unitary quantum evolution to the classical
percieved world of single well defined events. This probabilistic
limitation of the quantum theory has, since its conception, troubled
many physicists~\cite{bohr35nat,whitaker,scrod35,einsteinwir,everett,
  bohm52,bohmnature,grw86,hooftwitten}. For example Einstein stated in
a letter to Born in 1926: \emph{``I, at any rate, am convinced that
  {\em He} is not playing at dice.''} \cite{einsteinborn2}.
       
Essential insight to the measurement problem has been gained through
the decoherence program \cite{zurek03,schlosshauer,jooszeh}. Most
importantly, realistic macroscopic systems are never isolated, but
interacting with their environments. The interaction with the
environments forces the quantum wavefunction of macroscopic systems to
evolve into classical-like pointer
states~\cite{zurek03,schlosshauer,jooszeh}. Decoherence theory
explains the emergence of classical properties of systems coupled to
their surrounding environments. But the theory has nothing to say
about single outcomes!  Decoherence always refer to an {\em ensemble
  average} where the environmental degrees of freedom are traced out,
resulting in a reduced density matrix stating the probabilities to
find the system in different pointer states. There are no mechanism in
the Decoherence theory that chooses a single outcome among all the
possible pointer states. To many practitioners of quantum mechanics
this is not a problem, since they interprete the wavefunction to only
describe ensemble averages.  However, we here choose to interprete the
wavefunction as representing an individual system. With this
interpretation a state in superposition of outcomes {\bf a} {\em
  and\/} {\bf b\/} cannot suddenly evolve to an outcome of {\bf a}
{\em or\/} {\bf b}, even if decoherence is involved.

In this work, we show from the quantum evolution of pure states that
single outcomes of measurement events are possible as a consequence of
a phase transition in the measurement apparatus. The wavefunction of
the measurement apparatus is forced into classical-like pointer states
by the interaction with the environment. The destruction of
superpositions and, thereby, selection of a preferred pointer state is
caused by an infinitesimal non-linearity in the macroscopic
measurement apparatus. Which pointer state the wavefunction actually
collapses into depends on fine details of the initial conditions. The
probability for the wavefunction to collapse into a particular pointer
state is shown to be in close agreement with Born's rule. The
statistics for the outcomes depends on the size of the attraction
basins in the measurement apparatus which again is affected by the
initial state of the measured object.

To be more specific about our physical picture, consider first the
quantum dynamics of a closed Universe consisting of a linear
ferromagnet ($A$) embedded in a large environment ($E$). Assume that
the ferromagnet has an easy $z$-axis. Let the initial state of the
linear ferromagnet be a superposition between a state where nearly all
spins are pointing up and a state where nearly all spins are pointing
down, $|A_0 \rangle = | A_\nearrow \rangle + |A_\swarrow \rangle$, and
let the initial state of the environment be $|E_0 \rangle$. When the
ferromagnet interacts with the environment, the ferromagnet will
transfer its entropy and energy into the environment and relax towards
its two ground states, $ |A_\uparrow \rangle $ and $ |A_\downarrow
\rangle $, which turn out to be pointer states. But since the time
evolution is linear the dynamics of the distinct branches in the
superposition is completely independent of each other.  The
wavefunction of the whole system then becomes $|\Psi \rangle =
|A_\uparrow \rangle |E_\uparrow \rangle + |A_\downarrow \rangle
|E_\downarrow \rangle$ after some transient relaxation time. Already
here we see that interactions with an environment do not provide
single outcomes. {\em Quantum superpositions survive the interaction
  with environments}. The closed Universe evolves into independent
branches~\cite{everett}, here, a superposition of two pointer
states. There are no mechanisms that select a preferred evolution
branch and the ferromagnet remains in a superposition with zero
average magnetization. We clearly see the incapability of linear
quantum mechanics to describe everyday observations. Macroscopic
ferromagnets are not in superpositions, their magnetization have
preferred directions. We show below that a small non-linear term in
the ferromagnet may destroy the superpositions and select a preferred
direction.

Nonlinearity has been argued to exist intrinsically in quantum
mechanics~\cite{weinberg89}, for example at the Planck
scale~\cite{Svetlichny:ijotp05}. More importantly, macroscopic quantum
coherent systems are often effectively described by nonlinear
Schrödinger equations~\cite{pang}.  For example the dynamics of the
quantum coherent superconducting state is governed by the nonlinear
Ginzburg-Landau equation~\cite{Tinkham:Book75}. Another example is
Bose-Einstein condensates where the mesoscopic quantum coherent
wavefunction is governed by the nonlinear Gross-Pitaevskii
equation~\cite{BEC}. The fractional quantum Hall state may also be
described by an effective nonlinear equation~\cite{Zhang:prl89}.
% \bibitem{Zhang:prl99} S. C. Zhang, T. H. Hansson and S. Kilvelson,
%   Phys. Rev. Lett. {\bf 62}, 82 (1989). 
From the ``emergence'' point of view, these kind of nonlinearity may
naturally emerge as a consequence of the enormous number of
interacting particles involved in the macroscopic coherent state which
may have less symmetry compared to the underlying many-body
Hamiltonian~\cite{Anderson:Science72}. Finally, nonlinearity may also
rise naturally as a consequence of the interaction between two
fluctuating quantum fields in the mean-field limit, see
Appendix~\ref{App:energy}. In this paper we view nonlinearity as
modeling physical reality, not only a convenient (and necessary)
approximation.

A tiny nonlinearity will, to be shown below, force a macroscopic
ferromagnet embedded in an environment to chose a preferred direction,
consistent with experimental observations. For ferromagnets, one may
argue for a physically sound, nonlinear term by considering the
self-induced magnetic field created by the magnetization. Assuming
that the spins in a ferromagnet are carried by charged particles, then
in connection to the spin each particle also possess a small magnetic
moment, $g \mu_B \mathbf{S}$, where $\mathbf{S}$ is a dimensionless
spin operator, $g$ is the gyro magnetic factor and $\mu_B$ is the Bohr
magneton. The magnetic moments create a magnetic field which may be,
crudely, approximated to $\mathbf{B} = \mu_0 g \mu_B \langle \Psi |
\sum_{i \in A} \mathbf{S}_i | \Psi \rangle$ where $\mu_0$ is the
magnetic constant and $|\Psi \rangle$ is the conventional normalized
wavefunction of the system. Here, we have completely neglected spatial
variations, fluctuations and other internal degrees of freedom in the
magnetic field. We believe that those additional degrees of freedom in
$\boldsymbol{B}$ does not fundamentally modify the physical picture
presented below. It is important to note that we use the operator
$\langle \Psi |...| \Psi \rangle$ to denote a (inner) scalar product,
{\em no ensemble average is included in the process.}  The
self-induced magnetic field may again interact with the spins in the
ferromagnet giving rise to a nonlinear term in the Hamiltonian which
may, {\em at the mean field level}, be expressed as
\begin{eqnarray}
  H_{B} = - \mu ~ \langle \Psi | \sum_{i \in A} \mathbf{S}_i | 
  \Psi \rangle \cdot \sum_{j \in A} \mathbf{S}_j,
\label{HMagn}
\end{eqnarray}
where the parameter $\mu$ controls the nonlinearity and has the
dimension of energy.  This non-linear term will favor the pointer
state where the spins are parallel to the magnetic field and disfavor
all other pointer states. Here is the physical picture. {\em Each
  measurement apparatus has a set of attraction basins, related to the
  set of eigenvalues. Each attraction basin has a stable fixed point
  that is the pointer state}. Interactions with environments will
force all the parts of a superposition that reside in an attraction
basin to relax into the basin's stable fixed point, i.e. the basin's
pointer state. A self-induced magnetic field, $\mathbf{B}$, above a
certain threshold will single out a preferred attraction basin and
transform all other initially attraction basins into repulsive basins
and their stable fixed points into unstable fixed
points. Consequently, the measurement apparatus has now a single
unique attraction basin with its corresponding pointer state. Hence,
the ferromagnet will, in the presence of a self-induced magnetic field
together with an environment, evolve into a single unique pointer
state.

In the case of no coupling to an environment, there is no mechanism
for the spins in the ferromagnet to relax and the ferromagnet is then
forced to remain in a superposition regardless of nonlinearity. With
an environment, however, non-zero fluctuations in the magnetization
will create a magnetic field which, when strong enough, enhances
parallel spins and thereby increase the magnetization along that
direction. Thus, environment induced fluctuations may starts a
self-enhancing process which favors the pointer state with spins
parallel to the direction of the initial fluctuation. This process
forces the system to choose a preferred attraction basin. During this
process, the interactions with the environment also cause the
wavefunction to fall into the fixed point of the chosen attraction
basin, i.e. its pointer state. The probability for the wavefunction to
end up in a given pointer state is governed by the size of the
attraction basins, which for a ferromagnet alone is of course $50\%$
for ending as $|\Psi \rangle = |A_\uparrow \rangle |E_\uparrow
\rangle$ and $50\%$ for ending as $|\Psi \rangle = |A_\downarrow
\rangle |E_\downarrow \rangle$. We will discuss the the size of the
attraction basins in more details below. In summary, a non-linear term
in the Hamiltonian as Eq.~\eqref{HMagn} may somewhere along the time
evolution of the ferromagnet chose a preferred attraction basin. In
the same time, interactions with the environment will force the
wavefunction to fall into an unique classical-like pointer state which
is the stable fixed point of the chosen attraction basin. Which
attraction basin that is actually chosen depend on small details of
the initial state of the ferromagnet and the environment. An
infinitesimal small change in the initial state may force the
wavefunction to collapse in a entirely different pointer
state. However, the dynamics is completely deterministic.  Given the
same initial condition, the wavefunction will always choose the same
attraction basin and collapse into the same final pointer
state. \emph{In this sense, quantum mechanics is a deterministic
  theory}.

Nonlinearity in the ferromagnet can in principle be infinitesimally
small and still do its job in the selection of the preferred
attraction basin. The reason is as follows. For simplicity, consider a
ferromagnet in 2 dimensions with the Hamiltonian
\begin{eqnarray}
  H_I = -J \sum\limits_{\langle i,j \rangle} S_i^z S_j^z
\label{IsingHam}
\end{eqnarray}
where $\langle i,j \rangle$ denotes nearest neighboring spin-{\small
  \textonehalf} spins. This ferromagnet has two degenerate ground
states, $|\!  \uparrow \uparrow ... \uparrow \rangle$ and $|\!
\downarrow \downarrow ... \downarrow \rangle$ with the energies
$E_\uparrow = E_\downarrow = -JN_A/2$ where $N_A$ is the number of
spins in the ferromagnet. The ground states are separated by an energy
barrier with height $-JN_A/2$. A magnetic field given by
Eq.~\eqref{HMagn} will lift the degeneracy between $E_\uparrow$ and
$E_\downarrow$ by $\Delta E = E_\uparrow - E_\downarrow = \mu
N_A^2/4$. Note that this argument also applies for the case where the
magnetic field decays a as a function of distance from its sources
provided that the exchange interaction between the spins decays much
faster. Thus for large $N_A$, the energy from the initially
fluctuating magnetic field will not only define a global minimum, but
also eventually grow strong enough in order to completely remove the
local minimum connected to the anti-parallel ground state forcing both
branches of the wavefunction to fall into the same fixed point defined
by the global minimum. Therefore, an infinitesimally small nonlinear
parameter $\mu$ is sufficient to define a global minimum and remove
all other local minima for a macroscopic ferromagnet.

\begin{figure}[h]
  \begin{picture}(0,160)
%    \put(-70,0){\includegraphics[scale=0.4]{figura/Born.eps}}
    \put(-70,0){\includegraphics[scale=0.4]{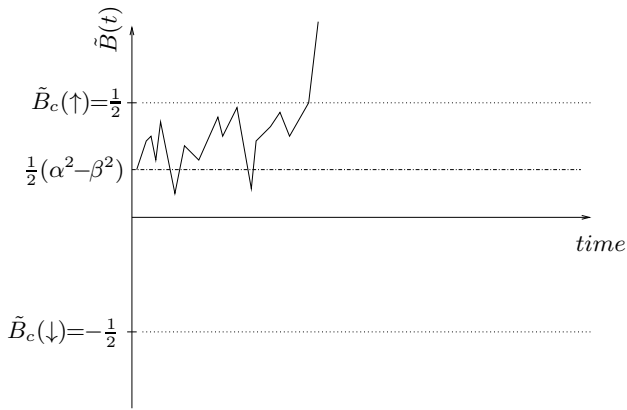}}
    \put(-82,135){\rotatebox{90}{$\tilde{B}(t)$}} %
    \put(100,60){$time$} %
    \put(-106,114){$\tilde{B}_c(\uparrow) \keq \frac{1}{2}$} %
    \put(-115,27){$\tilde{B}_c(\downarrow) \keq -\frac{1}{2}$} %
    \put(-109,88){$\frac{1}{2}(\alpha^2 \!\!-\!\! \beta^2)$} %
  \end{picture}
  \caption{Schematic figure of the dimensionless magnetic field as a
    function of time. Due to the initial state of the system spin
    $|S_{sys}(t \!=\! 0) \rangle = \alpha |\uparrow \rangle + \beta |
    \downarrow \rangle$, the dimensionless magnetic field start of its
    deterministic random walk from $\frac{1}{2}(\alpha^2 \!-\!
    \beta^2)$. When $\tilde{B}(t)$ reaches one of the two threshold
    values, non-linear effect will dominate over the fluctuations and
    force the ferromagnet to chose a direction.}
\label{fig:Born}
\end{figure}

We will from now on speak of the ferromagnet ($A$) as an measurement
apparatus, referring to the system as our measurement object.
Including a measurement object, that in our case is a system spin
$S_{sys}$ interacting with the ferromagnet, will alter the size of the
attraction basins of the ferromagnet. Here we provide an estimate for
the probabilities showing that this change of size provides statistics
in accordance with Born's rule. First, let us exclusively focus on the
case of a ferromagnet embedded in an environment with no measurement
object. For clarity we again restrict the discussion to an Ising
ferromagnet with spin-$\frac{1}{2}$ spins.  Define the dimensionless
magnetic field for the ferromagnet as $\tilde{B}(t) = \langle \Psi(t)
| \sum_{i \in A} S_i^z | \Psi(t) \rangle$, and let the initial
magnetization of the ferromagnet be zero, $\tilde{B}(t \keq 0) =
0$. Interactions with environments force $\tilde{B}(t)$ to
deterministically fluctuate back and forth around zero. Assume that
when $\tilde{B}$ reaches a threshold, $\pm \tilde{B}_c$, the nonlinear
term dominates over the fluctuations and starts a self enhancing
process which ends up in a macroscopic stable magnetization, see
Fig.~\ref{fig:Born}. A measurement apparatus that can measure the
direction of a single spin must be sensitive to a flip of a single
spin. Thus, $\tilde{B}_c \simeq \pm \frac{1}{2}$. In other words, when
the environment has flipped one spin in the ferromagnet, nonlinearity
will set in and force all other spins to align to that spin. Note that
this value of $B_c$ applies only for apparatus that are able to detect
the direction of a single spin-$\frac{1}{2}$ spin. If the initial
states of the ferromagnet and the environment are symmetric with
respect to the up and down pointer states, $\tilde{B}(t)$ will
fluctuate symmetrically around zero. Hence, the probability for that
$\tilde{B}(t)$ reaches $\tilde{B}_c$ first and the ferromagnet will
eventually choose the up state is $0.5$. Similarly, the probability
for that $\tilde{B}(t)$ reaches $-\tilde{B}_c$ first and the
ferromagnet will eventually end up in the down state is $0.5$, see
Fig.~\ref{fig:Born}. Note that the dynamic from a state with zero
magnetization into a state with macroscopic magnetization closely
resembles the dynamics of a phase transition. The state of the system
evolves from a state with the same symmetry as the underlying
Hamiltonian into a state with less (broken)
symmetry~\cite{Anderson:Science72}.

Now, add a system spin to be measured by a ferromagnet embedded in an
environment. Let the system spin be in the initial state
\begin{eqnarray}
  |S_{sys}(t \!=\! 0) \rangle = \alpha |\uparrow \rangle + \beta | \downarrow 
  \rangle
\end{eqnarray}
where $\alpha$ and $\beta$ are complex scalars. Let for simplicity the
initial state of the ferromagnet be in an antiferromagnetic
configuration
\begin{eqnarray}
  |A_0 \rangle= |\uparrow \downarrow \uparrow \downarrow ... 
  \uparrow \downarrow \rangle.
\end{eqnarray}
Of course, other states including those with superpositions which has
zero magnetization will do the job. After a short time the system spin
has interacted with the ferromagnet and flipped its spin leading to a
combined state where one spin in the ferromagnet is flipped, e.g
\begin{eqnarray}
  \Psi = \alpha | \downarrow \rangle |\uparrow \uparrow \uparrow 
  \downarrow ... \uparrow \downarrow \rangle + \beta | \uparrow \rangle 
  |\downarrow \downarrow \uparrow \downarrow ... \uparrow \downarrow \rangle.
\end{eqnarray}
The dimensionless magnetic field is now $\tilde{B} =
\frac{1}{2}(\alpha^2 - \beta^2)$ so the starting point of
$\tilde{B}(t)$ is no longer at zero but at $\frac{1}{2}(\alpha^2 -
\beta^2)$, see Fig.~\ref{fig:Born}. We see also that the size of the
``up'' attraction basin is $\frac{1}{2} + \frac{1}{2}(\alpha^2 -
\beta^2)$ while the size of the "down" attraction basin is
$\frac{1}{2} - \frac{1}{2}(\alpha^2 - \beta^2)$. The probability ratio
for $\tilde{B}$ to reach $\tilde{B}_c$ first versus $-\tilde{B}_c$ is
therefore, see Fig.~\ref{fig:Born},
\begin{eqnarray}
  \frac{P_\uparrow}{P_\downarrow} = \frac{\frac{1}{2} + 
    \frac{1}{2}(\alpha^2 - \beta^2)}
  {\frac{1}{2} -  \frac{1}{2}(\alpha^2 - \beta^2)} = 
  \frac{\alpha^2}{\beta^2}
\end{eqnarray}
In the last equality we have used the normalization condition
$\alpha^2 + \beta^2 = 1$. Thus, the probability for the ferromagnet to
end with a stable up state is $P_\uparrow = \alpha^2$. And, the
probability for the ferromagnet to end with a stable down state is
$P_\downarrow = \beta^2$, which are in accordance with Born's rule.

Our approach should not be confused with that of the dynamical
reduction program \cite{grw86,ghirardirev}, where the Schrödinger
equation is altered by inclusion of stochastic nonlinear terms in
order to achieve dynamical or spontaneous localisation of wave
packets. The basic idea behind the dynamical reduction program is that
the wavefunction randomly collapses all the time, where the rate of
collapse is related to the mass of the system.  This applies to all
particles whether isolated or interacting~\cite{grw86,ghirardirev}. In
our approach, an isolated system remains unperturbed until the
different systems start to interact. It is the asymmetry between the
size of the environment and the rest that causes a classical
appearance of the state of the measurement apparatus and the measured
object upon measurements.

Ref.~\cite{Hanson:quant-ph2000} considered a self-collapse mechanism
where the quantum wavefunction automatically exhibit dynamical
collapse, without any measurement apparatus nor environment, due to
nonlinearity originating from non-Abelian gauge fields. Our
description of a measurement process is completely different from
Ref.~\cite{Hanson:quant-ph2000}. We find that the dynamics of the
wavefunction strongly depend on the existence of a macroscopic
measurement apparatus embedded in an environment that is able to
absorb entropy and energy from the measurement apparatus.

Ref~\cite{wezel} studied a collapse of the quantum
wavefunction due to a spontaneous symmetry breaking process in the
measurement apparatus, an anti-ferromagnet in the thermodynamical
limit. To induce the collapse Ref~\cite{wezel} used a time
dependent, fluctuating, symmetry breaking, non-unitary, staggered
magnetic field with an undetermined origin.  A major difference
between our model and the model in Ref~\cite{wezel} is the
importance of the environment in the collapse of the wavefunction. We
studied the evolution of a pure wavefunction describing a closed
Universe.
%\bibitem{Wezel:arXiv08} J. van Wezel, arXiv:0804.3026v1, (2008).

Note that the results of this work by no means contradict the
predictions of quantum mechanics. Merely, the results shows that
quantum mechanics may be regarded as a complete theory which can
describe measurement processes and predicts single outcomes of
classical alike states without the collapse postulate. In other words,
the results show that quantum mechanics is a deterministic theory. The
price to pay is a tiny non-linearity effectively only in systems of
macroscopic sizes.

\section{Model}
To support the ideas discussed above in the introduction, we perform
numerical modelling of an idealized, but realistic, model of a quantum
measurement process. Our model is composed of a system spin,
$\boldsymbol{S}_{sys}$, being measured by a ferromagnet ($A$) embedded
in a spin-glass environment ($E$). The system spin is a
spin-$\frac{1}{2}$ spin. The measurement apparatus is a ferromagnet
consisting of a number $N_A=4$ or $N_A=8$ spin-$\frac{1}{2}$ spins. In
addition to the traditional spin-spin exchange couplings, the spins in
the ferromagnet also interact with its self-induced magnetic field,
giving rise to a weak non-linear term. Finally, the system spin and
the ferromagnet are embedded in an environment consist of a $N_E = 15$
spin-$\frac{1}{2}$ spins with random, frustrated, spin-glass
interactions.

\subsection{Hamiltonian}
Our model Hamiltonian may be written as
\begin{eqnarray}
  H&=&H_A+H_E+H_{AE}+H_{SA}+H_{SE}+H_B
\label{ham}
\\
H_A&=&-\sum\limits_{i,j\in A}
\sum\limits_{\alpha}J_{ij}^{\alpha}S_i^{\alpha}S_j^{\alpha} \nonumber \\
H_{AE}&=&\sum\limits_{i\in A}\sum\limits_{j\in E}
\sum\limits_{\alpha}\Delta_{ij}^{\alpha}S_i^{\alpha}I_j^{\alpha}\nonumber \\
H_E&=&\sum\limits_{i,j\in E}
\sum\limits_{\alpha}\Omega_{ij}^{\alpha}I_i^{\alpha}I_j^{\alpha} \nonumber\\
H_{SE}&=&\sum\limits_{i\in E} \sum\limits_{\alpha} \Theta_i^{\alpha} 
S_{sys}^{\alpha}I_i^{\alpha} \nonumber\\
H_{SA}&=&\sum\limits_{i\in A}\Gamma_i S_{sys}^{z}S_i^{z}\nonumber \\
H_{B}&=& - \mu  ~ \langle \Psi | \sum_{i \in
  A} S_i^z | \Psi \rangle ~\sum_{j \in A} S_j^z, \nonumber
\end{eqnarray}
where where $S_i^\alpha$, $I_i^\alpha$ and $S_{sys}^\alpha$ are
dimensionless spin-{\small \textonehalf} operators in the
ferromagnetic measurement apparatus, the spin-glass environment and
the system spin to be measured, respectively.  In Eq.~\eqref{ham}, the
index $\alpha = x,y,z$ runs over the three components of the spin
operators.  The exchange couplings $J_{ij}^{\alpha}$,
$\Omega_{ij}^{\alpha}$ and $\Delta_{ij}^{\alpha}$ control the
interaction between spins in the ferromagnet, the spin-glass
environment and between the spins in the ferromagnet and the
environment, respectively. Furthermore, $\Gamma_i$ controls the
interaction between the system spin and the apparatus while
$\Theta_i^\alpha$ controls the interaction between the system spin and
the environment. In more physical terms, the measurement apparatus is
modelled by a ferromagnet, $H_A$. A spin glass, $H_E$, serves the role
of an environment which absorbs the energy and entropy of the
ferromagnet allowing it to relax towards its ferromagnetic ground
states. The system to be measured is a single spin-$\frac{1}{2}$ spin,
$\boldsymbol{S}_{sys}$, which role is to tilt the up-down symmetry of
the ferromagnet.  The interaction between the ferromagnet and the spin
glass is modelled by $H_{AE}$. While $H_{SA}$ describes the
interaction between the system spin and the ferromagnet. To make the
measurement process more realistic, we have also a coupling between
the system spin and the environment, $H_{SE}$.

The Hamiltonian given by Eq.~\eqref{ham} without the non-linear term,
$H_B$, is linear and will not be able to conduct any measurement
process as discussed in the introduction and shown in more details
below. This linear part of the Hamiltonian can not choose a preferred
pointer states among the possible pointer states for the
ferromagnet. To model a ferromagnet that shows the behavior found in
experiments, we introduce a non-linear term, $H_B$, which describes
the interaction between the self-induced magnetic field and the spins
in the ferromagnet. Physically, the self-induced magnetic field ($B$)
originates from the magnetization of the ferromagnet, $B \propto M
\propto \sum_{i\in A} \langle \Psi | S_i^z | \Psi \rangle$. Again,
$|\Psi \rangle$ is the normalized wavefunction for the whole system,
i.e. closed Universe. We use the parameter $\mu$ to control the
non-linear self coupling between the magnetic field and the spins
$\boldsymbol{S}_i$ in the ferromagnet. In the macroscopic limit, an
infinitesimal small $\mu$ will force the ferromagnet to choose a
direction. Note that these kind of one-particle non-linear interaction
can also be argued to originate from linear terms like
$\boldsymbol{S}_i \cdot \boldsymbol{S}_j$ treated in a mean field
level $\boldsymbol{S}_i \cdot \boldsymbol{S}_j \rightarrow \langle
\boldsymbol{S}_i \rangle \cdot \boldsymbol{S}_j$. This kind of mean
field approximation has been used with tremendous success in other
macroscopic many-body quantum systems as
superconductivity~\cite{Tinkham:Book75}, superfluidity~\cite{BEC} and
fractional quantum hall
states~\cite{Tsui82,Laughlin83,Picciotto97}. There, the mean field
solution accurately describes the properties of complicated many-body
quantum wavefunctions in the macroscopic limit. Another source for the
nonlinear term is the interaction between $\Psi$ and a magnetic field
treated in the mean-field approximation, see
Appendix~\ref{App:energy}. Note that the Schrödinger equation used,
Eq.~\ref{ham}, is of Hamiltonian type even in the presence of a
nonlinear term. The energy is conserved during time evolution, see
Appendix \ref{App:energy}.

Now, to the parameters of the Hamiltonian, Eq.~\ref{ham}. We assume
$J^z_{ij} = J$ for nearest neighbors and $J/\sqrt{2}$ for next nearest
neighbors. The spin configuration of the ferromagnetic measurement
apparatus is rectangular (cubic) when composed of $4$ ($8$) spins,
respectively. For clarity we will, henceforth, use $J$ as our unit of
energy, i.e. $J=1$. All other {\em energies are measured with respect
  to $J$}. We use anisotropic coupling $J_{ij}^{x}=J_{ij}^{y}=\gamma
J_{ij}^{z} = \gamma$, where $\gamma\leq 1$, to make the ferromagnet
Ising-like. This is in order to reduce the infinite set of pointer
states for a Heisenberg ferromagnet to a set of two pointer states, up
and down, for a more transparent physical picture.  The exchange
couplings $\Omega_{ij}^{\alpha}$, $\Delta_{ij}^{\alpha}$ and
$\Theta_i^{\alpha}$ are uniform random numbers in the range
$[-\Omega,\Omega]$, $[-\Delta,\Delta]$ and $[-\Theta,\Theta]$, for all
$\alpha$. The coupling between the system spin to be measured and the
measurement apparatus $\Gamma_{i}=1$, for all $i \in A$.

A similar Hamiltonian without the system spin and the non-linear self
induced magnetic field $B$ has been studied in
Ref.~\cite{yuanmain}. There, they study, numerically, how the spin
glass (E) relaxes the ferromagnet (A)~\cite{yuanmain,yuanbath,
  yuangiant}. It was found that frustrated spin-glasses where very
effective with respect to relaxation and decoherence of the central
system, even for small numbers of spins ($N_E\approx10$) in the
environments. Spin glasses are therefore ideal as environments for
small spin systems.

\subsection{Preparation and numerical method}
This section describes the preparation of the initial state and the
procedure for the numerical time integration of the Schrödinger
equation.

First, we prepare an initial state
\begin{eqnarray}
  |\Psi(t=0) \rangle = |S_{sys}(t \!=\! 0) \rangle \otimes |A_0 \rangle 
  \otimes |E_0 \rangle. 
\label{InitState}
\end{eqnarray}
The system spin to be measured is prepared in a general superposition
\begin{equation}
  |S_{sys}(t \!=\! 0) \rangle = |\uparrow\rangle\cos{\theta}
  + |\downarrow \rangle\sin{\theta}e^{i\phi},
\label{preparation}
\end{equation}
where the phase angle $\phi\in[0,2\pi]$ is not so important for the
physical picture and henceforth set to 0. The ferromagnet is for
simplicity prepared in an antiferromagnetic state $|A_0\rangle = |
\uparrow \downarrow ... \uparrow \downarrow \rangle$. Of course, other
states with zero magnetization would do the same job.  The environment
is prepared in a state close to the spin-glass ground state
$|E_0\rangle$ computed by the Lanczos method \cite{hollenberg}.

Next, we iteratively time integrate the Schrödinger equation $i
\frac{d\Psi}{dt} = H \Psi$ using the finite difference method,
\begin{eqnarray}
  \Psi(t+dt) = \Psi(t) - i H \Psi dt,
\label{SchEq}
\end{eqnarray}
starting from the initial state, Eq.~\eqref{InitState}. The
Hamiltonian $H$ is given by Eq.~\eqref{ham}. For clarity, we have set
$\hbar=1$ in Eq~\eqref{SchEq} and throughout the article. Each
iterative step in Eq.~\eqref{SchEq} is found numerically by the
Chebyshev expansion method \cite{tal,zhang,dobrovitski}. The method is
based on a polynomial expansion of the propagator
$U(t,t_0)=e^{i(t-t_0)H}\approx J_0(t)+2\sum\limits_{k=1}^K
J_k(t)T_k(H)$, where $T_k(x)$ is the Chebyshev polynomial of the first
kind and $J_k(x)$ is the Bessel function of integer order $k$. The
Chebyshev moments $T_k(H)$ is computed by the recursive application of
the Hamiltonian operator $H$ on the wavefunction $|\Psi\rangle$.  We
have verified that the numerical method preserves the norm of the
wavefunction, $\langle \Psi(t) | \Psi(t)\rangle$ for all
$t$. Furthermore, the conserved energy of the closed universe
\begin{eqnarray}
  E_U = \langle \Psi | H - \frac{1}{2} H_B |\Psi \rangle
\label{E_tot}
\end{eqnarray}
has also been verified to be conserved in the numerical
calculations. A short derivation of this form of the conserved energy
for our model $E_U$ is given in Appendix~\ref{App:energy}.

The Hilbert space of the composite system $H =H_{S}\otimes H_{A}
\otimes H_{E}$ is of dimension $2^L$, where $L$ is the total number of
spins, system spin plus ferromagnet plus spin-glass. Present computing
power limits us to model systems where $L \le 24$.

The following parameters are fixed in all simulations: $\Delta = 0.3$,
$\Omega = 0.8$, $\Theta = 0.5$, $N_E = 15$. The remaining parameters:
$N_A = 4 \rightarrow \mu=12.0$ or $N_A = 8 \rightarrow \mu=6.0$. The
nonlinear parameter $\mu$ is varied such that the effect of maximum
magnetization on each spin in the ferromagnet is constant, $\mu
N_A=48.0$. This allows direct comparisons between the measurement
results from ferromagnets of different sizes. Since our closed
Universe is very small, the non-linear parameter had been set
relatively large to optimize the measurement properties of our tiny
measurement apparatus. In the macroscopic limit the non-linear
parameter may be infinitesimal small and still do the same job, as
argued for in the introduction.

\section{Results}
In this section we present the numerical results for the model given
by Eq.~\eqref{ham}. The results support our claim that the
wavefunction of the measurement apparatus, in the presence of a small
nonlinear term, is forced to choose one of the two classical pointer
states $|A_{\uparrow} \rangle$ or $|A_{\downarrow} \rangle$.  Without
the non-linear term, the ferromagnet remains in the superposition
regardless of the interaction with the environment. We also present
statistics for measurements with different initial configurations for
various $\theta$, and show that the presence of a measurement object
will alter the size of the attraction basins in the measurement
apparatus in a way such that the outcome is consistent with Born's
rule to the numerical precision.

\subsection{Time evolution}
It is difficult to extract useful information directly from the
complex many-body quantum state of the whole system which is, here, a
$2^{20}$ or $2^{24}$ dimensional complex vector. In order to
characterize the ferromagnet, we use the dimensionless magnetization
\begin{eqnarray}
  M(t) = \sum_{i\in A} \langle \Psi(t) | S_i^z | \Psi(t)\rangle
\end{eqnarray}
which determines the degree of ferromagnetic order in the measurement
apparatus. In addition, we use the exchange energy
\begin{eqnarray}
  E(t) = - \sum_{i,j\in A} J_{ij}\langle\Psi(t)|
  \boldsymbol{S}_i \cdot \boldsymbol{S}_j | \Psi(t)  \rangle
\end{eqnarray}
to characterize the relaxation of the ferromagnet. To characterize the
state of the system spin we use the expectation value
\begin{eqnarray}
  \langle S_{sys}^z(t) \rangle = \langle \Psi(t) | S_{sys}^z | \Psi(t) \rangle.
\end{eqnarray}
Again, note that we use the operator $\langle \Psi |...| \Psi \rangle$
to denote a (inner) scalar product, {\em no ensemble average is
  performed.}

\begin{figure}[htb]
  \begin{picture}(0,140)
    \put(-130,0){\includegraphics[width=0.5\textwidth]{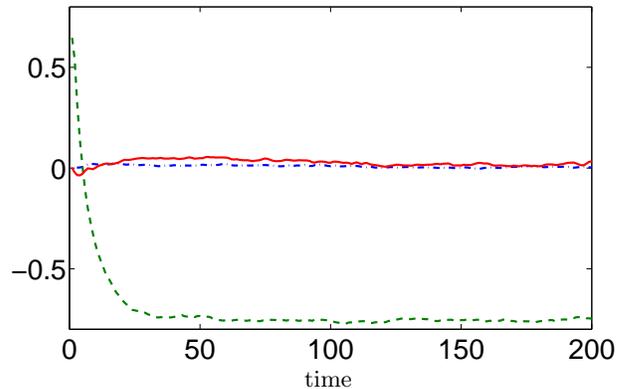}}
    \put(-8,-5){time} %
  \end{picture}
  \caption{Typical time evolution of the magnetization $M$ (solid) and
    the exchange energy $E$ (dashed) for a linear ferromagnet and the
    expectation value of the system spin (dashed dot). The initial
    state of the system spin is $|S_{sys}(t \!=\!  0) \rangle =
    \frac{1}{\sqrt{2}}(| \uparrow \rangle + | \downarrow
    \rangle)$. The number of spins in the ferromagnet is $N_A=4$.}
  \label{linear}
\end{figure}
Fig. \ref{linear} shows a typical time evolution of the magnetization
$M$ and the exchange energy $E$ characterizing the state of a linear
ferromagnet, i.e. $\mu=0$. The system spin is initially prepared in
the state $\left |S_{sys}(t \!=\! 0) \right\rangle
=\frac{1}{\sqrt{2}}(\left |\uparrow\right\rangle +\left|\downarrow
\right\rangle)$. We see that the exchange energy of the ferromagnet
decreases rapidly, as the ferromagnet relaxes from the
anti-ferromagnetic state towards its ferromagnetic ground states,
$|A_\uparrow \rangle$ and $|A_\downarrow \rangle$. Note that due to
the limited number of spins used in the ferromagnet, $N_A=4$, the
interaction with the large environment prevents the ferromagnet to
relax all the way down to the ferromagnetic ground states where
$E=-1.35$. We also see in Fig.~\ref{linear} that the magnetization $M$
shows small fluctuations, but no sustained net magnetization
characterizing the expected classical ground state. Fig.~\ref{linear}
also shows that the expectation value of the measured system spin is
zero, indicating that $\left|S_{sys} \right\rangle$ remains in a
superposition between $\left|\uparrow \right\rangle$ and
$\left|\downarrow \right\rangle$. A more careful examination of the
stationary state shows that the composite wavefunction of our closed
universe is still in a superposition. Thus, the magnetization
fluctuates around zero and the ferromagnet does not choose any
preferred pointer state. From Fig.~\ref{linear}, one may conclude that
linear ferromagnets are not able to conduct any measurement. The
superposition persists regardless of the interaction with the
environment. The time evolution
\begin{equation}
  (| \uparrow  \rangle+|\downarrow \rangle)
  |A_{0}\rangle|E_0\rangle 
  \rightarrow 
  | \uparrow  \rangle|A_{\uparrow}\rangle| 
    E_{\uparrow}\rangle+|\downarrow \rangle
  |A_{\downarrow}\rangle|E_{\downarrow}\rangle,
\label{linevolution}
\end{equation}
describes the dynamics in the linear case.  The initial state of the
system becomes entangled with the apparatus and the environment. The
two branches of Eq. \eqref{linevolution} evolve completely independent
of each other, and the initial superposition has through interaction
with its environment extended to our entire closed universe. Thus,
decoherence, in the sense of relaxation to the pointer states, alone
is not sufficient to remove the initial superposition. It should be
noted that the persistence of the superposition in this case is
entirely due to the exact linearity of the dynamical equation, and has
nothing to do with the size of the apparatus. It is straight forward
to show that superpositions survive any linear finite sized
ferromagnet, see Appendix \ref{App:lin}.

\begin{figure}[htb]
  \begin{picture}(0,150)
    \put(-130,0){\includegraphics[width=0.5\textwidth]{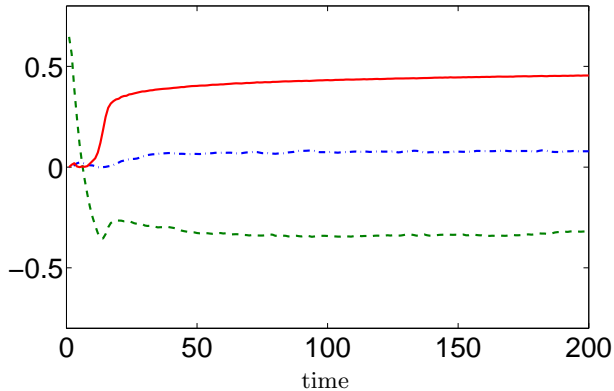}}
    \put(-8,-6){time} %
  \end{picture}
  \caption{A typical time evolution of the magnetization $M$ (solid)
    and the exchange energy $E$ (dashed) of a non-linear ferromagnet,
    and the expectation value $\langle S_{sys}^z\rangle$ (dashed dot).
    The system spin is initially prepared in the state $|S_{sys}(t
    \!=\!  0)\rangle = \frac{1}{\sqrt{2}} ( |\uparrow\rangle +
    |\downarrow\rangle )$. The number of spins in the ferromagnet is
    $N_A=4$.}
  \label{equalsuperpos}
\end{figure}
\begin{figure}[htb]
  \begin{picture}(0,150)
    \put(-130,0){\includegraphics[width=0.5\textwidth]{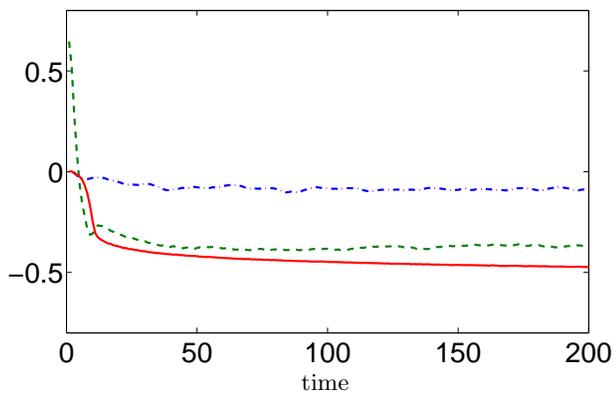}}
    \put(-8,-5){time} %
  \end{picture}
  \caption{Same as in Fig. \ref{equalsuperpos}. The only difference is
    the initial state of the environment.}
  \label{equalsuperpos2}
\end{figure}

Typical time evolutions of the magnetization $M$ of a non-linear
ferromagnet is shown in Figs.~\ref{equalsuperpos} and
\ref{equalsuperpos2}. After initial fluctuations due to the
environmental interaction, one of the two possible magnetization
directions is enhanced due to the feedback provided by interaction
with the self-induced magnetic field $B$. A ``phase transition'' in
the ferromagnet takes place, where the ferromagnet relaxes into the
pointer state of the attraction basin selected by the emerged magnetic
field. We have quoted the term ``phase transition'' here since our
tiny ferromagnets consist of only $4$ spins and phase transitions
happens in principle only in the thermodynamical limit. Rather, finite
size effects~\cite{Privman} are expected to dominate the behavior of
our small ferromagnets. This can be seen in the actual values for
$M(t)$, $E(t)$ and $S_{sys}(t)$. They are all significantly less than
the values expected in the thermodynamic limit which are:
$M=0.5*N_A=2.0$, $E=-1.35$ and $S_{sys}=0.5$. However, the stationary
values of $M$, $E$ and $S_{sys}$ are stabilized well beyond the noise
and one may clearly conclude that a measurement has been conducted by
an admittedly rather poor measurement apparatus.

In more physical terms, when the magnetization reaches a critical
value, the non-linear term sets in and turns all attraction basins,
except the chosen one, into repulsive basins. From this point there is
no turning back, all superpositions in the wavefunction of the closed
Universe are forced to evolve such that the state of the ferromagnet
becomes the unique pointer state chosen by the magnetic field. Thus,
as a result of the measurement dynamics the spin expectation value of
our system $\langle S_{sys}\rangle$ will follow the direction of the
established magnetization $M$. The state of the measurement object has
been driven into the state corresponding to the final outcome of the
measurement apparatus. The final result is in correspondence to the
result obtained by the collapse postulate of orthodox quantum theory
where an initial state $(|\uparrow\rangle+|\downarrow\rangle)\otimes
|A_0\rangle$ is instantaneously and discontinuously collapsed upon
measurement to either $|\uparrow\rangle\otimes |A_{\uparrow}\rangle$
or $|\downarrow\rangle\otimes |A_{\downarrow}\rangle$.  We might
therefore conclude that a quantum measurement of a single event has
been accomplished. The total wavefunction is
\begin{eqnarray} |\Psi \rangle = |\uparrow \rangle |A_\uparrow \rangle
  |\tilde{E}_\uparrow \rangle
\end{eqnarray}
after a measurement where the up pointer state is chosen. Note that
the final state of the environment $|\tilde{E}_\uparrow \rangle$ still
contains superpositions inherited from the initial superposition of
the measured spin.

The only difference in the initial state $|\Psi\rangle$ of the
composite system between Fig. \ref{equalsuperpos} and
Fig. \ref{equalsuperpos2} is the different initial state of the
environment. Note that the different initial states of the environment
are completely equivalent, they all belong to the set of degenerate
ground states of the spin glass. We see that a small difference in the
initial state of the environment does in this case alter the dynamics
of the measurement process completely. In this example the slight
altering of the initial state resulted in a different final pointer
state of the measurement apparatus. It is this sensitivity to the
initial state that gives rise to the indoctrinated randomness in
orthodox quantum mechanics, and to Born's rule.

\subsection{Statistics}
According to Born's rule \cite{born}, in an ideal quantum mechanical
measurement the probability for our measurement apparatus to end in
the state of positive magnetization $|A_{\uparrow}\rangle$,
when the object is prepared in the state $|S_{sys}(t \!=\! 0) \rangle$ of
Eq.~\eqref{preparation} is
\begin{equation}
  P_\uparrow(\theta) = \cos^2{\theta}.
\label{prob2}
\end{equation}
Correspondingly, the probability for ending in the negative
magnetization state is
\begin{equation}
  P_\downarrow(\theta) = \sin^2{\theta}.
\label{prob3}
\end{equation}
To obtain measurement statistics, we run a number of $96$ independent
simulations for each choice of $\theta$. For each simulation, a
different spin glass ground state is used as the initial state for the
environment.

\begin{figure}[htb]
  \begin{picture}(0,150)
    \put(-130,0){\includegraphics[width=0.5\textwidth]{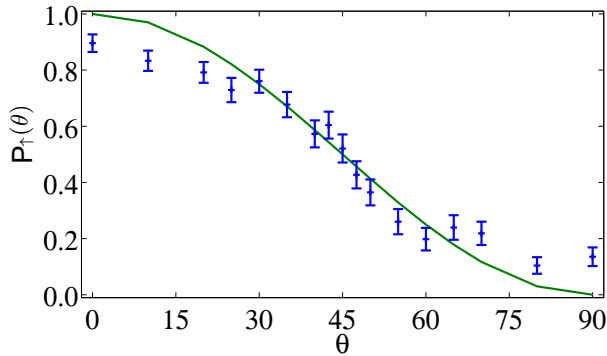}}
    \put(-124,81){\rotatebox{90}{$_\uparrow(\theta)$}}
  \end{picture}
  \caption{Probability for a ferromagnet of 4 spins to end in an up
    pointer state as a function of the angle $\theta$ of the initial
    system spin state $|S_{sys}(t \!=\! 0) \rangle = | \uparrow
    \rangle \cos{\theta} + | \downarrow \rangle \sin{\theta}$. Solid
    line shows ideal statistics according to Born's rule.}
  \label{stat4}
\end{figure}
The probability for the measurement apparatus and the system spin to
end up as a function of $\theta$ for a ferromagnet consist of $N_A=4$
spins is shown in Fig.~\ref{stat4}. Here, the non-linear parameter is
$\mu=12.0$. We have verified that $P_\uparrow(\theta) +
P_\downarrow(\theta) = 1$. We see in Fig.~\ref{stat4} that the
probabilities of obtaining the pointer state $|\uparrow \rangle
|A_{\uparrow}\rangle$ as the outcome of the measurement process
resemble Born's rule. The discrepancies at small and large $\theta$
are due to the huge finite size effect in the tiny measurement
apparatus. The subsystem of system spin plus apparatus are so small,
that the interaction with the large environment affects the
measurement result. Needless to say, a 4-spins ferromagnet is far from
a perfect measurement apparatus. It is, for us, surprising that this
tiny ferromagnet actually provides results so close to Born's
rule. One reason may be due to the up-down symmetry of the ferromagnet
and of the environment. Since the subsystem of system spin plus
ferromagnet is so small, the ``noise'' from the environment dominates
and the probability for ending up is close to $50\%$ regardless of the
initial state of the system spin. Thus, $P_\uparrow(\theta=45)=0.5$ is
guaranteed by symmetry.

\begin{figure}[htb]
  \begin{picture}(0,150)
    \put(-130,0){\includegraphics[width=0.5\textwidth]{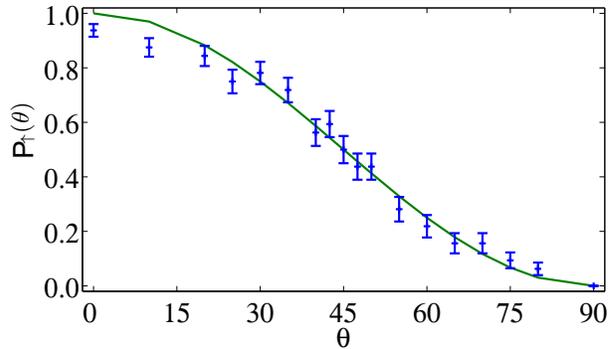}}
    \put(-124,81){\rotatebox{90}{$_\uparrow(\theta)$}}
  \end{picture}
  \caption{Probability for a ferromagnet of 8 spins to end in an up
    pointer state as a function of the angle $\theta$ of the initial
    system spin state $|S_{sys}(t \!=\! 0) \rangle = | \uparrow
    \rangle \cos{\theta} + | \downarrow \rangle \sin{\theta}$. Solid
    line shows ideal statistics according to Born's rule.}
  \label{stat8}
\end{figure}
To show that larger ferromagnets will actually provide better
measurement results, we have also carried out simulations with a
system with a $N_A=8$ spins ferromagnet. Here, the non-linear
parameter is $\mu=6.0$. Recall that, this value of $\mu$ is chosen to
conserve the effect of maximum magnetization on each spin in the
ferromagnet, $\mu N_A = 48$, such that a direct comparison with the
$N_A=4$ case is possible. All other parameters are identical as
above. As seen in Fig.~\ref{stat8}, the probability for the
ferromagnet to end in an up state follows Born's rule more closely
compared to the $N_A = 4$ spins ferromagnet. This is expected, since
the finite size effect is smaller here. By comparing the cases $N_A=4
\rightarrow \mu=12.0$ and $N_A=8 \rightarrow \mu=6.0$, we see that the
nonlinear parameter $\mu$ decreases for increasing size of the
ferromagnet. Thus, for a macroscopic system the non-linear parameter
$\mu$ may be infinitesimally small and still do its job: selecting a
preferred attraction basin and consequently a pointer state.

Obviously the obtained statistics is highly dependent on the choice of
parameters, for example on the strength of the nonlinear interaction
$\mu$. We have chosen the parameters of our model in such a way that
our measurement apparatus is optimized for measuring the $z$-component
of a single spin.

\section{Conclusion}
We show from a model study of a simple closed Universe that definite
outcomes of quantum measurements can emerge continuously from pure
quantum evolution.  Each measurement apparatus has a set of attraction
basins corresponding to the set of eigenvalues. The selection of a
preferred attraction basin in a measurement process is caused by an
infinitesimally small nonlinearity in the measurement
apparatus. Interaction with the environment then forces all
superpositions of the measured object and the measurement apparatus to
fall into the same fixed point (pointer state) of the preferred
attraction basin.  The dynamics of this measurement process which
strongly resembles the dynamics through phase transitions is entirely
deterministic!  Given the same initial condition the wavefunction will
always fall into the same pointer state. However, small changes in the
initial state, for example in the environment, may cause the
wavefunction to fall into a completely different pointer state. Thus,
the uncontrollable degrees of freedom in the environments account for
the intrinsic statistical behavior of the quantum measurement
process. The probability for falling into a certain pointer state is
govern by the size of the attraction basins of the measurement
apparatus and is shown to be in close agreement with Born's rule.

How the dynamics of non-local experiments, e.g. the EPR
experiment~\cite{epr}, will be affected by non-linear measurement
apparatus is an interesting future study.

To conclude, we have presented a model where the probabilistic
behaviour of Quantum mechanics has exactly the same origin as the
random outcomes of a rolling dice, i.e.\ its sensitive behaviour of
uncontrollable initial conditions. Thus, even though neither {\em
  He\/} nor {\em Quantum Mechanics\/} are playing at dice, we
physicists still have to.

\begin{acknowledgments}
  We thank S. Girvin, J.-P. Morten, A. Sudb{\o}, H.-J. Bergli for
  stimulating discussions and specially A. Brataas for making this
  project possible. This work has been supported by the Norwegian
  research council through the grants no. 162742/V00,
  167498/V30. Computing power and resources have been provided by the
  Notur project.
\end{acknowledgments}

\appendix

%\section{Appendixes}
\section{Conservation of energy}
\label{App:energy}

In this appendix, we derive the expression for the conserved energy
for our model, Eq.~\eqref{E_tot}. The non-linear term is shown to
introduce an extra term in the expression for the conserved energy in
addition to the standard average of the Hamiltonian. The appendix also
shows how a non-linear term as the one used in our model may naturally
arise from the interaction between two different fields.

For clarity, we consider a simple quantum field model having the basic
properties of our model, Eq.~\eqref{ham}. The Lagrangian of our
quantum field model with Einstein sum-convention over repeated indices
assumed can be written as
\begin{equation}
  L=i\Psi_a^{\star} \dot{\Psi}_a - \Psi_a^{\star} H_{ab}\Psi_b - 
  \Phi_{\alpha} \Psi_a^{\star}\Gamma_{ab}^\alpha \Psi_b
  +\frac{1}{2}\epsilon \dot{\Phi}_\alpha \dot{\Phi}_\alpha-V(\Phi).
  \label{lagrangian}
\end{equation} 
Here, $\Psi$ is a complex field representing the wavefunction and
$H_{ab}$ is the Hamiltonian for $\Psi$ where $a$ and $b$ are
multi-indices which may include spin number, spin direction and also
continuous variables like space coordinated etc. In
Eq.~\eqref{lagrangian}, $\Phi$ is a scalar field which may be
associated with for example a harmonic oscillator with mass $\epsilon$
or in certain limit the electromagnetic field. The multi-index
$\alpha$ denoting all the variables necessary to describe the field
$\Phi$ which lives in a potential $V(\Phi)$.

From the Euler-Lagrange equation
\begin{eqnarray}
  \partial_t \left(\frac{\partial L}{\partial(\partial_t \Psi_a^{\star})}\right)
  -\frac{\partial L}{\partial \Psi_a^{\star}} &=&0  \\
  \partial_t \left(\frac{\partial L}{\partial(\partial_t \Phi_{\alpha})}\right)
  -\frac{\partial L}{\partial \Phi_{\alpha}} &=&0,
  \label{eulerlagrange1}
\end{eqnarray}
we deduce the dynamical equations
\begin{eqnarray}
  i\dot{\Psi}_a &=& H_{ab}\Psi_b+\Phi_{\alpha}\Gamma_{ab}^{\alpha} \Psi_b 
\label{EuLa1}
\\
  \epsilon \ddot{\Phi}_{\alpha} +\frac{\partial}{\partial\Phi_{\alpha}} 
  V(\Phi)&=& -\Psi_{a}^{\star}\Gamma_{ab}^{\alpha} \Psi_b
\label{EuLa2}.
\end{eqnarray}
Without the coupling between the fields when $\Gamma_{ab}^\alpha = 0$,
Eq.~\eqref{EuLa1} becomes the conventional Schrödinger equation for
$\Psi$ in the first quantized language. Furthermore, Eq.~\eqref{EuLa2}
take the form of Newton's second law for the field $\Phi$.

To deduce the conserved energy we now assume that the field $\Phi$ is,
similar to the electromagnetic field, mass-less i.e $\epsilon = 0$ and
that the potential is of the form
\begin{equation}
  V(\Phi)=\frac{1}{2}\Phi_\alpha \Phi_\alpha
  \label{potential}
\end{equation}
corresponding to the potential energy for the electromagnetic field
$\frac{1}{2}B^2$.
Equation \eqref{EuLa2} is then simplified to the constraint equation
\begin{equation}
  \Phi_{\alpha}=-\Psi_a^{\star}\Gamma_{ab}^{\alpha} \Psi_b,
  \label{constraint}
\end{equation} 
which inserted into the Lagrangian Eq. \eqref{lagrangian} give the
effective Lagrangian for the field $\Psi$
\begin{eqnarray}
  L' =  i\Psi_a^{\star} \dot{\Psi}_a- \Psi_a^{\star} H_{ab}\Psi_b
  -\frac{1}{2} \Phi_\alpha \Psi_a^{\star}\Gamma_{ab}^{\alpha} \Psi_b.
  \label{lagrangian2}
\end{eqnarray} 
We can now deduce the conserved energy
\begin{eqnarray}
  E = \Psi_a^{\star} H_{ab}\Psi_b
  +\frac{1}{2} \Phi_\alpha \Psi_a^{\star}\Gamma_{ab}^{\alpha} \Psi_b.
\label{conservedenergy}
\end{eqnarray} 

The conserved energy in our model, Eq.~\eqref{ham}, is then
\begin{eqnarray}
  E_U = \langle \Psi | H - \frac{1}{2} H_B |\Psi \rangle
\end{eqnarray}
if one relate the interaction term $-\Phi_{\alpha}
\Psi_a^{\star}\Gamma_{ab}^\alpha \Psi_b$ in Eq.~\eqref{lagrangian}
with our non-linear term $H_B = - \mu (\sum\limits_{i\in A} \langle
S_i^z\rangle ) \sum\limits_{j\in A} S_j^z$ and $\Phi$ with
$\sum\limits_{i\in A} \langle S_i^z\rangle$. Thus, we see that the
nonlinear term $H_B$ gives rise to an additional term
$-\frac{1}{2}H_B$ in the expression for the conserved energy compares
to the conventional $E=\langle H \rangle$. 

More importantly, one see that the interaction between the fields
$\Phi$ and $\Psi$ give rise to a nonlinear term for the dynamic of
$\Psi$ in the mean field limit, i.e when $\Phi \rightarrow \langle
\Phi \rangle$.

\section{Linear time evolution}
\label{App:lin}
We will in this section show that superpositions survive interaction
with arbitrary macroscopic measurement apparatus as long as the time
evolution is linear.  Consider a system to be measured, initially
prepared in the superposition
$\left|\uparrow\right\rangle+\left|\downarrow\right\rangle$.  The
system is being measured by a measurement apparatus $A$, interacting
with an environment $E$.  We can write the initial state of the
composite system as
\begin{equation}
\label{eq:initial}
\left|\Psi\right\rangle=(\left|\uparrow\right\rangle+
\left|\downarrow\right\rangle)\otimes\left|A_{0}\right\rangle\otimes\left|
  E_0\right\rangle 
\end{equation}

According to linear quantum mechanics, time evolution is governed by
the time evolution operator $U(t,t_0)$, acting linearly on any quantum
state
\begin{equation}
\label{eq:U}
U(t,t_0)(\Psi_A+\Psi_B)=U(t,t_0)\Psi_A+U(t,t_0)\Psi_B.
\end{equation}
Thus the time evolution of branch $A$ of the wavefunction is
completely independent of branch $B$.

When applied to a quantum measurement linear quantum evolution lead to
the von Neumann chain of infinite regress \cite{neumann}
\begin{equation}
  \left(\left| \uparrow \right \rangle+\left|\downarrow \right\rangle\right)
  \left|A_{0}\right\rangle\left|E_0\right\rangle 
  \stackrel{t}{\rightarrow} 
  \left|\uparrow \right \rangle\left|A_{\uparrow}\right\rangle
  \left|E_{\uparrow}\right\rangle+\left|\downarrow \right\rangle
  \left|A_{\downarrow}\right\rangle\left|E_{\downarrow}\right\rangle.
\label{eq:linevolution}
\end{equation}
Since the environment ($E$) can in principle denote every degree of
freedom in the universe, including observers, we see that linear time
evolution result in the initial superposition of the system being
measured eventually extending to the entire universe.  This time
evolution (i.e linear quantum mechanics without the collapse
postulate), naturally lead to the Everett relative state
interpretation \cite{everett}. Where the universe continuously split
in distinct independent branches.

% Produces the bibliography via BibTeX.
%\bibliography{spinbib}

\end{document}